\documentclass[conference]{IEEEtran}
\IEEEoverridecommandlockouts
\usepackage{cite}
\usepackage{amsmath,amssymb,amsfonts}
\usepackage{autobreak}
\usepackage{algorithmic}
\usepackage{graphicx}
\usepackage{textcomp}
\usepackage{floatrow}
\usepackage{xcolor}
\usepackage{bm}
\usepackage{subfigure}
\usepackage{color}
\usepackage{times}
\usepackage{soul}
\usepackage{url}
\usepackage[hidelinks]{hyperref}
\usepackage[utf8]{inputenc}
\usepackage[small]{caption}
\usepackage{stfloats}
\usepackage{amsthm}
\usepackage{booktabs}
\usepackage{algorithm}
\usepackage{multirow}
\allowdisplaybreaks

\def\BibTeX{{\rm B\kern-.05em{\sc i\kern-.025em b}\kern-.08em
    T\kern-.1667em\lower.7ex\hbox{E}\kern-.125emX}}
\begin{document}
\makeatletter 
\makeatother

\newtheorem{MDP}{Definition}

\title{\LARGE{OSS Mentor: A framework for improving developers' contributions via deep reinforcement learning}}

\author{
\IEEEauthorblockN{Jiakuan Fan}
\IEEEauthorblockA{\textit{School of Data Science and Engineering} \\
\textit{East China Normal University}\\
Shanghai, China \\
jkfan@stu.ecnu.edu.cn}
\and
\IEEEauthorblockN{Haoyue Wang}
\IEEEauthorblockA{\textit{School of Data Science and Engineering} \\
\textit{East China Normal University}\\
Shanghai, China \\
51195100024@stu.ecnu.edu.cn}
\and

\IEEEauthorblockN{Wei Wang}
\IEEEauthorblockA{\textit{School of Data Science and Engineering} \\
\textit{East China Normal University}\\
Shanghai, China \\
wwang@dase.ecnu.edu.cn}
\and

\IEEEauthorblockN{Ming Gao}
\IEEEauthorblockA{\textit{School of Data Science and Engineering} \\
\textit{East China Normal University}\\
Shanghai, China \\
mgao@dase.ecnu.edu.cn}
\and 
\IEEEauthorblockN{Shengyu Zhao}
\IEEEauthorblockA{\textit{college of Electronical and Information Engineering} \\
\textit{Tongji University}\\
Shanghai, China \\
frank\_zsy@tongji.edu.cn}

}

\maketitle

\begin{abstract}
In open source project governance, there has been a lot of concern about how to measure developers' contributions. However, extremely sparse work has focused on enabling developers to improve their contributions, while it is significant and valuable. In this paper, we introduce a deep reinforcement learning framework named Open Source Software(OSS) Mentor, which can be trained from empirical knowledge and then adaptively help developers improve their contributions. Extensive experiments demonstrate that OSS Mentor significantly outperforms excellent experimental results. Moreover, it is the first time that the presented framework explores deep reinforcement learning techniques to manage open source software, which enables us to design a more robust framework to improve developers' contributions.
\end{abstract}


\section{Introduction}

In recent years, open source software represented by Apache has achieved great success\cite{1}. Developers with different development experiences in different regions gather together spontaneously because of their interests, prestige, and employment needs \cite{2}. Developers of open source software generally contribute to the project by sharing experience \cite{3}, debugging code \cite{4}, and submitting functional patches \cite{5}. Many studies have always used project contribution as an important indicator to evaluate the status of developers \cite{6}.\par 

There are many use cases in which we need to compare and recognize different developers’ contributions. While traditional value-based software engineering \cite{i7,i8,i9} focuses on creating economic value as a way to prioritize resource allocation and scheduling, other measurements of value may be more relevant in some of the use cases. One example is that instructors need a tool with which to evaluate individual students’ code contributions to group projects (besides non-code contributions). Such measurement of code contributions has nothing to do with economic returns. As a second example, an engineering manager may need a quantitative measurement of team members’ performance. Additionally, for open-source software projects, developers’ contributions heavily influence collaboration, coordination, and leadership \cite{i10,i11}.\par

Therefore, modeling all the data of the top contributors of the project and analyzing the correlation between all actions quantitatively so that they can provide insights for improving the  contributions of developers, which is the main research significance of this topic. Increasing developers' contributions help software engineering project managers to set up project teams based on developers' profiles, thereby improving the productivity and code quality of the teams' development. As far as we know, there are no guidances on how to improve the contributions of developers in open source projects, and more on how to participate in open source projects. Because there are many ways for developers to contribute, and it is difficult to get a general guide.\par

At present, when developers want to participate in open source projects, the more common guidances are contribution guidelines. The contribution guidelines are textual documentation files, which embody a software project’s contribution process and  document the contribution expectations of project maintainers. 
However,  there has yet to be an exploration of what contribution guidelines contain and whether projects adhere to the workflows they prescribe. Currently, almost no one is doing anything similar to improve developers' contributions in open source software, while it is significant and valuable.\par

In this paper, we first define the model for the contribution, which reflects the mutual important relationships between actions over time, with considering all the possible actions (both coded and non-coded) from the perspective of the whole project. Further, we propose the Open Source Software(OSS) Mentor framework to help developers maximize their contributions by translating the actual problem into a reinforcement learning problem. In addition, we significantly improve the performance of the algorithm by enhancing the utilization of parameters during training. The main contributions of our paper:\par
1. We propose a data-based contribution evaluation model, which can dynamically measure developers' contributions based on changes in data.\par
2. We address the challenge of how to improve developers' contribution, which is an extremely rare and significant work at present.\par
3. It is the first time that the presented framework explores deep reinforcement learning techniques to manage open source software, which enables us to design a more robust framework to improve developers' contributions.\par
4. We have performed extensive experiments, proving the remarkable success of our proposed framework. \par
The main structure of the paper is as follows: In chapter 2 we focuses on our proposed framework OSS Mentor. In Chapter 3 we validate our model. In Chapter 4 we present the related work, and the discussion and summary sections are presented in Chapters 5 and 6.

\section{OSS Mentor Framework}
In this section, firstly, we give a framework for contribution assessment and show the overall architectural diagram of the proposed model. Immediately afterward, we quantify the essential elements of reinforcement learning in the context of practical problems. After that, we illustrate the algorithmic flow of the model based on the previous foundation. Finally, we describe the training process of the model.\par

\subsection{Overview}
$\bm{Definition \;of \;contribution.}$  \; Previously, the work on measuring developers' contribution was basically at the visual level, and the method of quantification was basically to directly count the number of issues, issue comment, PR, PR comment, etc., and to quantify the contribution by adding up the empirical empowerment\cite{report}. But this method is unreliable because there is no analysis of the project's data to obtain results that conform to objective laws, and it does not reflect the characteristics of the respective projects and changes over time. However, in our work, we first introduced the concept of entropy to measure how recognizable the developer's actions are to the developer, and then analyzed project-wide data to determine the weighting relationship between actions from an objective perspective. The entropy is calculated as shown in the formula:

\begin{equation}
H(X)=-\sum_{i=1} ^{n}P(x_{i})logP(x_{i})	\label{entropy}
\end{equation}
\quad $P(x_{i})$ represents the probability of the event $x_{i}$. In information theory, entropy represents the degree of discrete information, and the higher the degree of discrete, the greater the entropy and the greater the amount of information represented. So entropy is greatest with the discrete degree presenting an average distribution. In our work, action events actually executed by developers in open source projects are selected, and the discrete degree of action is measured by calculating the entropy value of the action event. If the higher the entropy value of H(X) on the $i$-th action dimension, the greater the amount of information, then the less discernible the $i$-th action is to the developer, which means that everyone is more inclined to perform it. Next, we use the entropy method to determine the degree of importance of each action in an open source project.\par

However, a prerequisite for information entropy is the assumption of independence between actions. And in practice, because of the inter-information problem between actions (e.g. there is a strong correlation between issue and issue comment), it does not directly satisfy the entropy-weighted computational system. To solve the inter-information problem, we replace information entropy with conditional entropy. The formula is shown in the figure:
\begin{equation}
H_{i}(Y|X)=-\sum_{x\in X,y\in Y} p_{i}(x,y)log \frac{p_{i}(x,y)}{p_{i}(x)}	\label{conditional entropy}
\end{equation}
\quad We overcome the problem of the assumption that actions in open source projects are not independent between each other by introducing conditional probabilities, e.g., issue and issue comment are not independent between two actions. With the above method, we can calculate the weight vector $W_{i}$ for each action dimension of the project. Finally, we get the calculation of the contribution:\par
\begin{equation}
C_{(i,k)}=\sum _{t=1}^T W_{(i,k)} * A_{(i,k)}^{t}	\label{contribution}
\end{equation}

where $A_{(i,k)} ^{t}$ denotes how many times the action is executed on the $t$-th step of the $k$-th episode of the $i$-th project. $W_{(i,k)}$ is computed from the conditional entropy model, and it has been normalized, which means $\sum_{t=1}^{T}W_{t}=1$. \par

Our work on the determination of the weights is extremely significant. First of all, the determination of the weights is based entirely on data, unlike previous work\cite{report} which is artificially determined through expert experience. Second, and most importantly, the weights are dynamic. That is, changes in project data over time and changes in project status, among other factors, can cause the weights to be updated. Therefore, we define weights that reflect not only the mutual importance relationships between actions over time, but also the important relationships that are specific to the actions between different projects.\par

\begin{figure}[htbp]
\centerline{\includegraphics[width=3.3in]{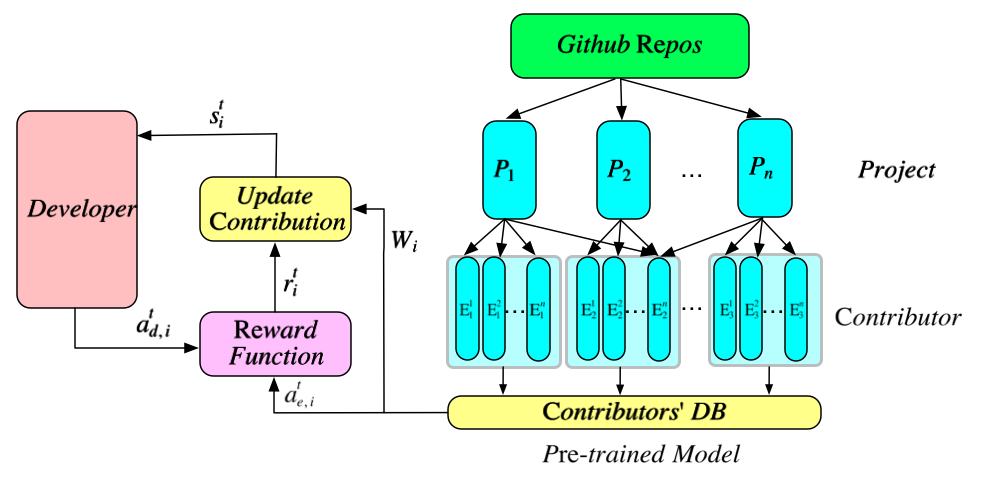}}
\caption{The overall framework of the proposed model. The environment section is a pre-trained model that uses the contribution quantification method (Equation \ref{contribution}) to pre-train.}
\label{fig}
\end{figure}

$\bm{Proposed\;model.}$ \quad This is the first time that deep reinforcement learning has been explored in the field of open source project governance. The goal of the model is to maximize the cumulative contribution of the developer after multiple executions of the action. The detailed flow of the model is given in Figure 1. First, the environment section is a pre-trained model that uses the contribution quantification method (Equation \ref{contribution}) to pre-train the weight vector $W_{i}$, and contains the contributors' action dataset $E_{i}=\left\{ e_{i}^{1},e_{i}^{2},e_{i}^{3},\cdots,e_{i}^{n} \right\}$. Developer as an Agent selects action $a_{d,i}^{t}$ in state $s_{i}^{t}$ according to its own policy. At this point, $a_{d,i}^{t}$ matches the sequence of actions $a_{e,i}^{t}$ with the corresponding contributor that matches the current state of the developer's ability from a project $P_{i}$ in the environment, and get the reward $r_{i}^{t}$ for performing the action by calculating the similarity between the two actions $a_{d,i}^{t}$ and $a_{e,i}^{t}$. After that, the contribution $C_{i}^{t}$ is refreshed by $W_{i}$ which is from the pre-trained model, then the state moves to the next state. Continuously, the strategy is updated.\par

This translates the question of how to improve the developer's contribution in open source software into a reinforcement learning question to maximize the contribution of the project. In addition, our proposed framework OSS Mentor can also recommend adaptive difficulty levels for different states of the developer's ability level, which will be combined with the current state, behavior pattern, and cumulative contribution to match the next most appropriate behavior pattern for the developer's ability.\par

\subsection{Elements formalization}
$\bm{Action.}$\;We define Action $A=[a_{1},a_{2},a_{3},\cdots,a_{n}]$ , which represents a collection of actions that developers can perform in an open source project, any element of the collection has the specific form: $a_{i}^{t}=[N_{a_{i1}}^{t}, N_{a_{i2}}^{t},..., N_{a_{in}}^{t}] $ represents the number of times the action was performed on each dimension in month $t$ for the $i$-th project. The indexes of each dimension of the vector indicate the actions that the developer can perform, such as open issue, issue comment, close issue, open pull request, pull request comment, etc. The value corresponding to the vector's dimensional index stands for the number of times the action has been performed in the current month.\par

$\bm{State.}$\;We define state $s_{i}^{t}=[e_{a_{i1}}^{t},e_{a_{i2}}^{t},...,e_{a_{im}}^{t},c_{a_{i}}^{t}]$, where the state is composed of the short-term feature $[e_{a_{i1}}^{t},e_{a_{i2}}^{t},...,e_{a_{im}}^{t}]$ of the reaction's immediate state and the long-term feature $c_{a_{i}}^{t}$ of the reaction's cumulative contribution. The status is divided into two main parts. The first part includes the expected value $\frac{1}{N}\sum e_{ai}^{t}$ for all contributor data for the same contribution of the developer performs. This part reflects the approximate optimal value of short-term features. The second part reflects the long-term feature of the developer's cumulative contribution and is used to describe the current proficiency  of the developer.\par

$\bm{Reward.}$\; The designed reward directly embodies the goal we want to optimize, and each reward earned indicates how well or badly the action is being executed, so designing a good reward function is very important to be able to successfully achieve the goal in a reinforcement learning problem. The goal is to maximize the developer's contribution, and Equation \ref{contribution} shows the calculation of the contribution. There is an obvious problem that the more times a developer executes, the more contributions he gets. But that shouldn't be the case, if every action performed by the developer is not similar to the contributor's action, then this execution of the action is supposed to be bad. Instead, it's a good action. Hence, it is very critical that reward is designed with the ability to measure the similarity between developer actions and contributor actions in a given contribution scenario. With this consideration in mind, we define the reward function:
\begin{equation}
r_{t} = W * A_{t}* exp(-\lambda \left\|(a_{d}^{t}-a_{e}^{t})*W_{t}\right\|_{2}^{2})	\label{reward}
\end{equation}
where $exp(-\lambda \left\|(a_{d}^{t}-a_{e}^{t})*W_{t}\right\|_{2}^{2})$ represents the similarity of developer's action $a_{d}^{t}$ and contributor's action $a_{e}^{t}$, $\lambda=\frac{1}{2\sigma^{2}}$, $\sigma$ is the control the width of the similarity domain between $a_{d}^{t}$ and $a_{e}^{t}$. $\left\|(a_{d}^{t}-a_{e}^{t})*W_{t}\right\|_{2}^{2}$ represents the Euclidean distance between the developers' actions and the contributors' actions, whereby different actions represent different weights, hence the weight vector is used to increase the impact of heavily weighted actions and decrease the impact of less heavily weighted actions. \par

To prevent the one-sidedness of measuring contribution only by the product of quantity and weight, such as increasing contribution only by improving the value $A_{i} ^{t}$, we added the penalty $\alpha_{i}^{t}$ to measure the similarity between the actions performed by the developer and the contributor. That is, $C_{i}$ is only as big as $A_{i} ^{t}$ and $\alpha_{i}^{t}$ are at the same time.\par

$\bm{Environment.}$ \quad We use the pre-trained model as an environment to interact with the agent. Data information for a global project is stored in the GitHub database, containing all event information defined by the GitHub API. For the projects to be tested, we extracted the higher-ranked contributors among them and used their historical behavioral data as a contributor pool, expecting to fulfill the general pattern learned through the contributor's pool by giving developers constant recommendations of actions that fit their own. For each action taken by the developer, the environment matches the behavior data of each contributor in the developer's current state in accordance with the project's contributor database data, what's more, it calculates the average behavior pattern as the environment's output for all the contributor data, finally, it calculates actions of developer and contributor to get the reward.\par

$\bm{Objective \; Function.}$ \quad Our goal is to maximize the developer's contribution. In combination with Equation \ref{reward}, we define the objective function as:
\begin{equation}
J_{t} ^{\theta^{k}}(\theta)=\sum_{(s_{t},a_{t})} min(\pi_{t}(\theta) A^{\theta^{k}}_{t}, clip(\pi_{t}(\theta), 1-\epsilon, 1+\epsilon)A^{\theta^{k}}_{t})	\label{ppo}
\end{equation}
\quad where $\pi_{t}(\theta)=\frac{p_{\theta}(a_{t}|s_{t})}{p_{\theta^{k}}(a_{t}|s_{t})}$, it represents the proportion of importance sampling between the two distributions $p_{\theta}(a_{t}|s_{t})$ and $p_{\theta^{k}}(a_{t}|s_{t})$. $A^{\theta^{k}}_{t}$ is an advantage function of the parameter $\theta^{k}$. $\epsilon$ is a hyper-parameter, $\epsilon \in [0,1]$.


\begin{algorithm}[h]
    \caption{OSS Mentor}
    \begin{algorithmic}[1]
    \STATE Pre-train open source software project by Eq. \ref{conditional entropy}
    \STATE Input: Contributors' data $e_{i}={e_{i}^{1},e_{i}^{2},...,e_{i}^{l}}$, $W_{i}$ obtained from the pre-trained model;
    \STATE Initialize: critic network $Q(s,a|\Theta^{Q})$, actor network \newline $\mu(s|\Theta^{\mu})$ with weights $\Theta^{Q},\Theta^{\mu}$
    \FOR{each $i \in [1,M]$}
    \STATE Receive initial observation state $s_{0}$;
    \FOR{each $t \in [1,T]$}
    \STATE select developer's action $a_{d}^{t}$ according to the policy and exploration noise;
    \STATE execute $a_{d}^{t}$, observe new state $s_{t+1}$, observe contributors' action $a_{e}^{t}$;
    \STATE compute similarity of ($a_{d}^{t}$,$a_{e}^{t}$), and contribution $c_{t}$ by Eq. \ref{reward};
    \IF {t \% batch-size == 0}
    \STATE update actor network parameter $\Theta^{\mu}$ by Eq. \ref{ppo};
    \STATE update critic network $\Theta^{Q}$;
    \ENDIF
    \ENDFOR
    \ENDFOR

    \end{algorithmic}
    \end{algorithm}

\subsection{Algorithm process}
There have been two difficult issues in open source project governance, the first of which is how to quantify the contributions of developers. We innovatively introduced the concept of information entropy to measure the discernment of the action, and then solved the problem of calculating the degree of importance between the actions by the entropy weight method. In response to the problem of independence assumptions that do not satisfy the entropy law, we solve the problem of conditional independence assumptions by calculating conditional entropy, finally systematically and well-defined formulas for measuring developers' contribution is proposed.\par

Having solved how to quantify developers' contributions, we come to the second question of how to maximize the contribution. Before that, related work had largely focused on simply how to visualize and quantify contribution, but very little research had addressed how to improve it. And we think the most important thing is how to help developers make decisions to improve their ability to participate in open source projects. We abstract the practical requirements of how to enhance developer contribution in open source project governance into a reinforcement learning problem and try to maximize contribution in this way. The specific algorithm flow is shown in Algorithm1 OSS Mentor.\par

\subsection{Model training}
First, we extracted the trajectories of contributors from the GitHub, and used the pre-trained model to obtain the weights of each dimension of the action vector. Then, we proposed OSS Mentor, which allows us to optimize the parameters by achieving the batch-size in each training round instead of optimizing it after the previous training round like PPO. This allows better control over the direction of optimization of parameters, and experiments have shown that our results have been greatly improved.\par

\section{Experiment}
\quad In order to verify whether our proposed OSS Mentor framework can help developers to improve their contributions, we conducted a substantial number of experiments. Based on the empirical evaluation, we aim to answer the following research questions.\par
\textbf{RQ1} How does OSS Mentor perform compared with the baseline methods in improving contributions?\par
\textbf{RQ2} How does the key hyper-parameter affect the performance of OSS Mentor?\par
\textbf{RQ3} Is the framework more significant than original algorithm?\par
\textbf{RQ4} Does OSS Mentor's strategy still work if the developer's state is disturbed by outside interference?\par
In what follows, we first introduce the experimental settings, and then answer the above research questions in turn. Furthermore, we perform two case studies to illustrate how OSS Mentor works.\par

\subsection{Dataset}
Based on X-lab report\cite{report}, we picked the top 10 active projects from GitHub, which are guaranteed to be as diverse as possible in the types and the fields.\par
\textbf{ansible/ansible}: Ansible is a radically simple IT automation system. \par
\textbf{MicrosoftDocs/azure-docs}: Azure-docs is the open source documentation of Microsoft Azure.\par
\textbf{DefinitelyTyped/DefinitelyTyped}: DefinitelyTyped is for high quality TypeScript type definitions.\par
\textbf{flutter/flutter}: Flutter is Google's SDK for crafting beautiful, fast user experiences for mobile, web and desktop from a single codebase.\par
\textbf{elastic/kibana}: Kibana is the window into the Elastic Stack. Specifically, it's a browser-based analytics and search dashboard for Elasticsearch.\par
\textbf{kubernetes/kubernetes}: Kubernetes is an open source system for managing containerized applications across multiple hosts. \par
\textbf{facebook/react-native}: React Native brings React's declarative UI framework to IOS and Android.\par
\textbf{tensorflow/tensorflow}: TensorFlow is an end-to-end open source platform for machine learning. \par
\textbf{microsoft/TypeScript}: TypeScript is a JavaScript that compiles to clean JavaScript output.  \par
\textbf{microsoft/vscode}: Visual Studio Code is a distribution of the Code-OSS repository with Microsoft specific customizations released under a traditional Microsoft product license.\par 

We collected these GitHub projects and users via the GitHub API. All the projects collected spanning from the beginning of projects  to December 2019. According to the number of developers' commits, we extracted the behavior trajectories of the top 120 developers in each project. \par

\subsection{Baselines}
We compare our proposed approach OSS Mentor with the previous excellent algorithm DDPG, \cite{ddpg} which learns both Q functions and strategies, uses offline data and Bellman's equation to learn the Q function and uses the Q function to learn strategy, suitable for continuous and high-dimensional spatial situations. GAIL \cite{gail} uses the opposite idea of traditional reinforcement learning, by a generative adversarial network  to learn the reward function from the contributor's data firstly, and then the forward training, suitable for situations where the reward function cannot be determined or the reward is sparse.\par

\begin{figure*}[htbp]
\setlength{\abovecaptionskip}{1pt} 
\setlength{\belowcaptionskip}{-1pt}
\centering
\centerline{\includegraphics[width=7in]{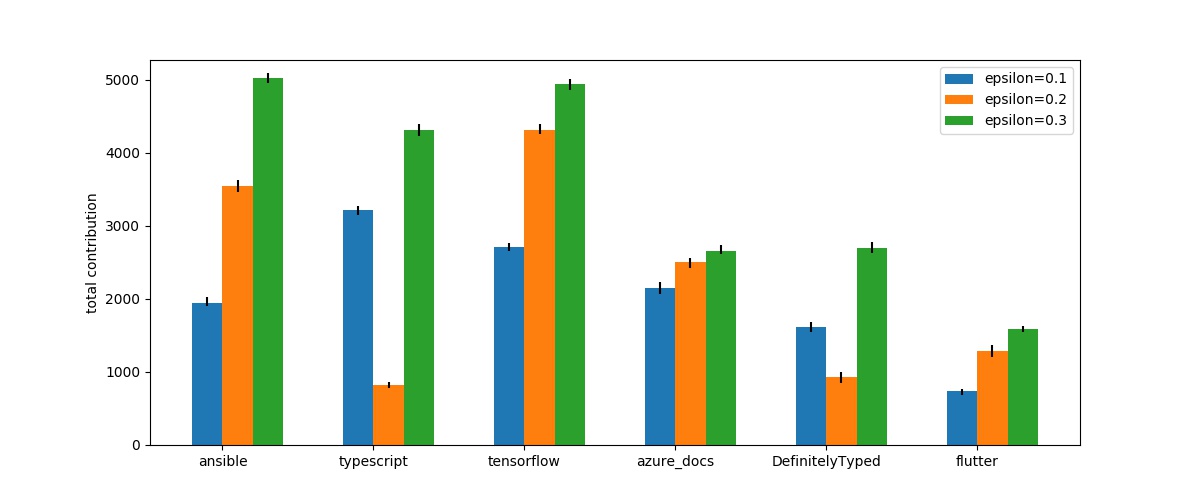}}
\caption{Control group experiment with hyperparameters. We set up 3 groups of control experiments to observe what the average contribution of each step is in the case where $\epsilon$ is 0.1, 0.2, and 0.3 . The result of each project is the value after averaging multiple times. It is clear that OSS Mentor achieves the best results when $\epsilon$ is 0.3. }
\vspace{-0.05cm}  
\setlength{\abovecaptionskip}{-0cm}   
\setlength{\belowcaptionskip}{-0.01cm}   
\label{fig:compare_fig1}
\end{figure*}

\subsection{Training details}
$\bm{K-Degree\;Relaxation.}$   \quad  We adopt the imitation learning mindset and use the extracted trajectories of higher contributors as a contributor pool, expecting to learn generally applicable laws to help the average developer continuously improve their contributions. So there is inevitably a problem with imitation learning, where the state distribution of the contributor is likely not to cover the state distribution of all developers, and this results in some extreme state developers being unable to learn. To solve this problem, we innovatively proposed the K-degree relaxation method to match contributor data. When a developer performs a specific action, the contribution represented by that action is matched to the actions performed by the contributors in the contributor pool with the same contribution. However, the match takes not an exact match but a fuzzy match, with each match contributing to the upper and lower K-degree space. When the contributor data cannot be matched in the K-degree space, a recursive mechanism is triggered, then the K-degree space is amplified outward until the relevant data are matched.\par

For the setup of the actor network, we constructed a network of 64 neurons, using the activation function ReLU. For the mean and variance of the network output, we used the activation function tanh,softplus, respectively.\par

For the reward function, we are finalizing the results by measuring the similarity between the developer's action and the contributor's action, and we have designed both cosine similarity and Gaussian core radial basis functions to measure this. However, the cosine similarity are optimized only for the direction between the vectors, and the length of the vectors, i.e., the number of actions performed, often converges to an extreme state that causes the strategy to fail, so we modified the Gaussian kernel radial basis function to measure the similarity very well.\par

\subsection{Parameters Setting}
\quad We construct two networks, Actor and Critic, the Actor network has a middle layer, and the mean and variance of the actions are computed by the network, respectively, the actions come from a random sample of a normal distribution constructed by the mean and variance. We have normalized the actions and states during the training process so that the model can train quickly and converge better. The learning rate of both actor and critic is set to 0.01, and the parameters of both actor and critic networks are updated after every 10 steps. For the CLIP mechanism, the upper and lower bounds were set to [1-$\epsilon$, 1+$\epsilon$]($\epsilon$=0.3). We truncated the probability of the old strategy and set the minimum value to 1e-5, in order to prevent the old strategy from not solving the gradient properly due to a value of 0.\par

\begin{figure*}[htbp]
\setlength{\abovecaptionskip}{1pt} 
\setlength{\belowcaptionskip}{-1pt}
\centering
\centerline{\includegraphics[width=7in]{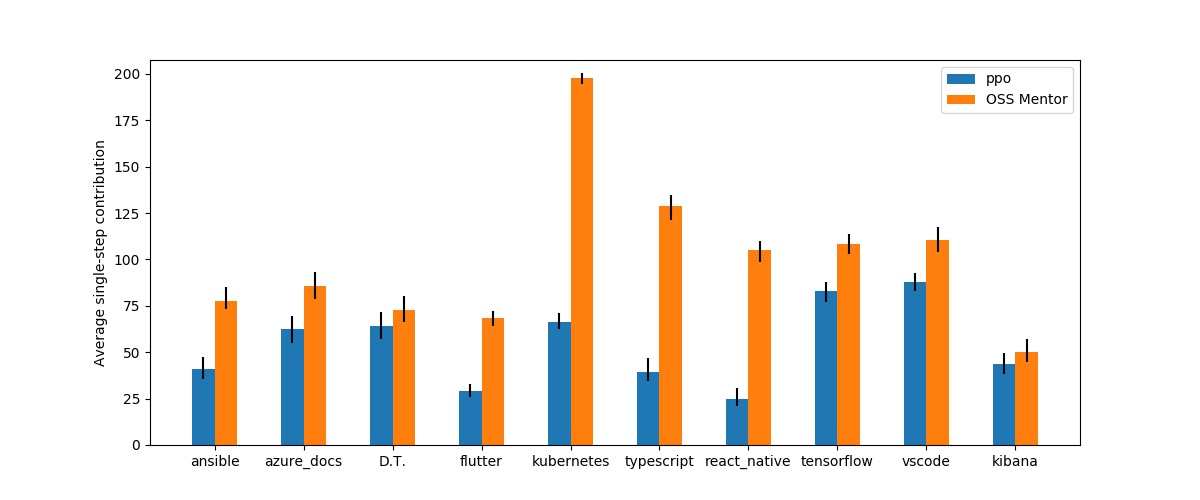}}
\caption{Comparative results of PPO and OSS Mentor. After algorithm training, we performed real-world tests on 10 projects on the two algorithms. From the selected screenshots, it can be seen that our OSS Mentor performs best in each project. In particular, OSS Mentor achieves more than three times performance of PPO in react-native open source project.}
\vspace{-0.05cm}  
\setlength{\abovecaptionskip}{-0cm}   
\setlength{\belowcaptionskip}{-0.01cm}   
\label{fig:compare_fig2}
\end{figure*}


\subsection{Experimental results}
$\bm{OSS\,Mentor\; VS \;Baseline(RQ1).}$  \quad To investigate the comparison between our proposed OSS Mentor performance and baseline performance, we performed this experiment. As shown in Table1, for each project of 10, we trained the GAIL, DDPG and OSS Mentor algorithms to find the optimal strategy to maximize the contribution. For each project we randomly selected the trajectories of 10 developers and experimented 3 times, taking the average of the 3 results as the final result to reduce the error caused by one experiment. From the data in the table, It is easy to see that our OSS Mentor demonstrated a tremendous advantage on the vast majority of projects. In addition, we also compared the developer's real track and got the results shown in the REAL column of the table. We think that the reason for this result is that our OSS Mentor is better able to address the learning step size choose the problem so that it can be optimized in a better direction every time. It is clear that OSS Mentor not only gives much better results than baseline, but also exhibits a huge advantage over real-life scenarios. \par

\begin{table}
\centering
\caption{Average single-step contribution of different algorithms}
\label{tab1}
\begin{tabular}{|c|c|c|c|c|}\hline
\multirow{2}*{Projects} &DDPG 				&GAIL 				&REAL			&OSS Mentor\\
\cline{2-5}
~ 		&						\multicolumn{4}{|c|}{contribution} 		 \\\hline
ansible 						&70.38				&14.80				&33.43			&$\bm{77.81}$			\\\hline
azure docs 						&85.68				&19.00				&14.18			&$\bm{85.90}$			\\\hline
D.T. 			 				&49.84				&13.89				&10.13			&$\bm{72.38}$			\\\hline
flutter 						&3.53				&13.18				&47.90			&$\bm{68.31}$			\\\hline
kubernetes 						&8.46				&20.46				&107.05			&$\bm{197.57}$			\\\hline
typescript 						&100.50				&12.88				&3.85			&$\bm{128.67}$			\\\hline
react native 					&12.97				&20.30				&15.62			&$\bm{104.85}$			\\\hline
tensorflow 						&10.44				&20.98				&26.09			&$\bm{108.33}$			\\\hline
vscode 							&16.23				&20.98				&11.16			&$\bm{110.48}$			\\\hline
kibana 							&89.05				&15.30				&16.01			&49.65	        \\\hline

\end{tabular}
\end{table}

$\bm{OSS\,Mentor\; with \;hyperparameter(RQ2).}$  \quad We investigate the impact of the hyper parameters $\epsilon$, which plays a crucial role to control the clip range. As shown in Fig 2. Each project is the result of averaging multiple tests. We can easily observe that as the value of the hyper-parameter becomes larger, the final contribution increases accordingly, except two projects. When $\epsilon$ is 0.3, it performs better than the others on all test projects. This is due to the factor that the strategy has a large clip range. Therefore, when the value of the hyper-parameter is 0.3, the algorithm will obtain  better performance.\par

$\bm{OSS \, Mentor \; VS \; PPO(RQ3).}$ \quad We designed this experiment to verify whether OSS Mentor has the better performance than the original model PPO. Figure 3 presents the results of the two comparisons, which is the average result of multiple experiments. From the selected screenshots, it can be seen that our OSS Mentor performs best in each project. In particular, OSS Mentor achieves more than three times the performance of PPO in react-native open source project. The reason for the result is that the parameter updating of the PPO algorithm occurs only after each round of training, which reduces the utilization of parameters during training. Our proposed OSS Mentor will update the parameters every batch-size length in each training round, thus retaining more information about the parameters, making the algorithm better. Therefore, our proposed OSS Mentor has better performance than the original algorithm.\par

\begin{figure}[htbp]
\setlength{\abovecaptionskip}{1pt} 
\setlength{\belowcaptionskip}{-1pt}
\centering
\centerline{\includegraphics[width=3.5in]{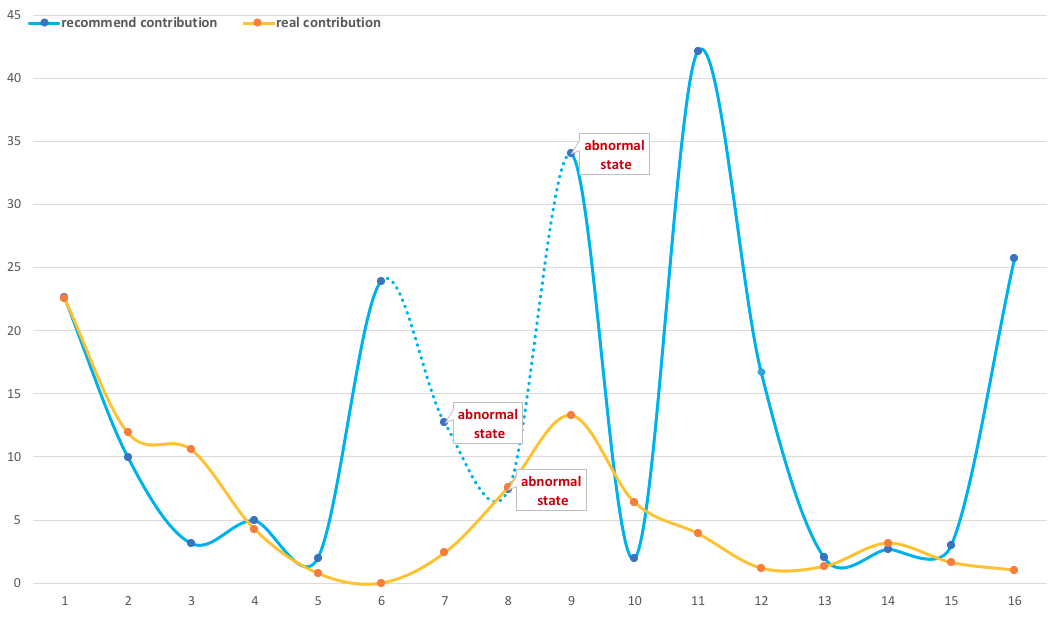}}
\caption{Intervention experiment. we interfere with the contributor's state in month 7,8,9 which is forced back to its initial value. The dashed line shows the contributions obtained under the OSS Mentor strategy after the state was disturbed.  We can see that the disturbed strategy still maintains a strong advantage over the true contribution.}
\vspace{-0.05cm}  
\setlength{\abovecaptionskip}{-0cm}   
\setlength{\belowcaptionskip}{-0.01cm}   
\label{fig:compare_fig4}
\end{figure}

$\bm{Intervention\; experiment(RQ4).}$  \quad To confirm that our proposed OSS Mentor has better algorithmic robustness, we designed an intervention experiment. The results are shown in Figure 4. We randomly selected a contributor's true trajectory and plotted his true contribution and OSS Mentor strategy change. Then, we interfere with the contributor's state which is forced back to its initial value in months 7, 8, and 9. The dashed line shows the contributions obtained under the OSS Mentor strategy after the state was disturbed. We can see that the disturbed strategy still maintains a strong advantage over the true contribution, and consistents with the contributing change characteristic -wavelike undulating rise- illustrated by the case study experiment. The reason why it works so well is that for each contributor's state we record its historical cumulative contribution value. Even when the state is destroyed, the strategy still recommends the most appropriate path based on the historical contribution. This ensures strong robustness of the algorithm.

$\bm{Case \;study.}$ \quad To show the effect of OSS Mentor in more detail, we randomly select a contributor's real trajectory in 18 months. Following his initial state, we use OSS Mentor to recommend strategies for the contributor. There is a common feature in both the true and recommended trajectories, where the contributor's contribution shows wavy ups and downs, but it is perfectly true that the spikes in contributions get higher and higher during the change. On the one hand, the contributor does not work intensely for many months, always taking short breaks. On the other hand, the maximum he can contribute is slowly increasing as he becomes more deeply involved in the project. The graph shows that the contributor did not contribute at all for six months after the fourth month of implementation, which is clearly not the best path. In OSS Mentor's strategy, a better strategy was adopted six months after month 4, taking into account the contributors' rest and work for intervals to boost their own contribution. The dotted line in Figure 5 shows the OSS Mentor's recommended method for better contribution enhancement.\par

\begin{figure}[htbp]
\setlength{\abovecaptionskip}{1pt} 
\setlength{\belowcaptionskip}{-1pt}
\centering
\centerline{\includegraphics[width=3.5in]{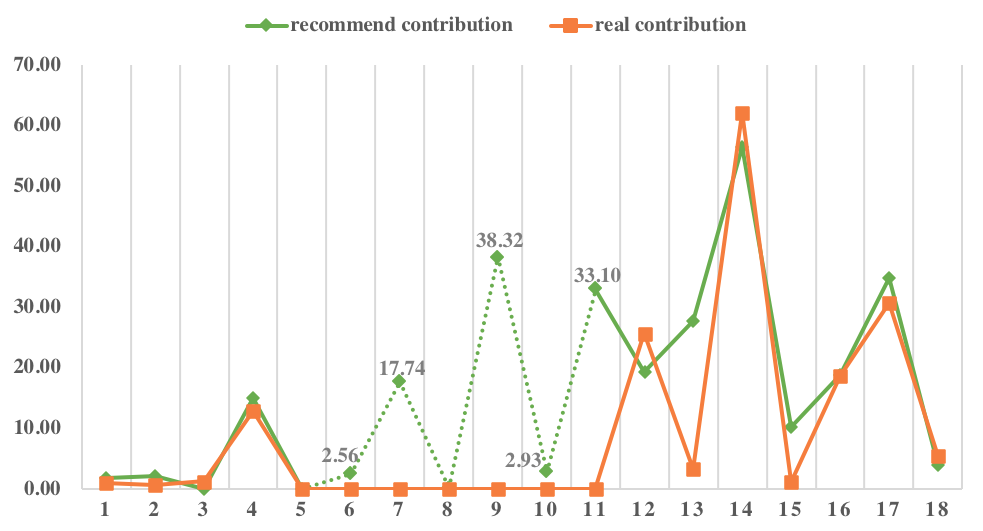}}
\caption{Case Study I: The horizontal coordinate represents the 18 months and the vertical coordinate represents the contribution received for each month. In OSS Mentor's strategy, a better strategy was adopted six months after month 4. The dotted line shows the our recommended method for better contribution enhancement.}
\vspace{-0.05cm}  
\setlength{\abovecaptionskip}{-0cm}   
\setlength{\belowcaptionskip}{-0.01cm}   
\label{fig:compare_fig5}
\end{figure}

\quad Another case study as shown in Figure 6 demonstrates that how the contribution of OSS Mentor and the baselines changes in detail on all projects, while the REAL represents the real changes of the 10 contributors. Experimental results are obtained by averaging multiple times. Due to space constraints, we have selected one of the 10 projects to illustrate, but the other projects are similar to this one. All of these projects, OSS Mentor consistently produces a higher contribution than others including the actual contributions performed by the contributors. In addition, it is clear that the  stability of our model is also better than the others, showing a steady upward trend, unlike DDPG and GAIL, which show negative optimization trends or oscillatory fluctuations. This means that our method has better stability and performance in every time.\par

\section{Related work}
\quad In the traditional software development model, the initial measure of developers' contribution is the number of words \cite{7}. The most classic measure is the number of lines of code written by the developer (Lines of Code), which is also the most simple way. LOC is the accumulation of the number of modified lines of all commits of each developer. There are 2 main code measurement methods, physical code line (physical SLOC, LOC) and logical code line measurement (logical SLOC, LLOC)\cite{8}. \par
Gousions et al. \cite{9}summarized nearly 30 metrics from data sources such as code warehouses, mailing lists and forums, defect databases, Wikis, etc. These metrics include both code and non-code content. They classified these indicators with positive and negative contributions according to their impact, and proposed a more complex measurement system. However, it was not fully used in the specific research process, nor did it study the importance of each indicator and whether it can be used to measure the contribution of developers.

\begin{figure}[htbp]
\setlength{\abovecaptionskip}{1pt} 
\setlength{\belowcaptionskip}{-1pt}
\centering
\centerline{\includegraphics[width=3.5in]{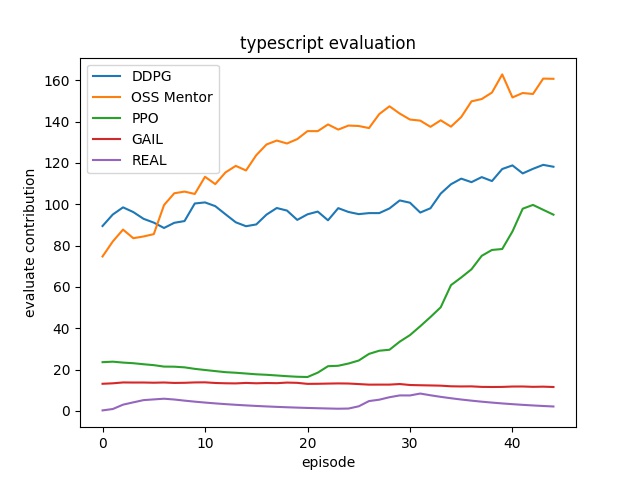}}
\caption{Case Study II: It represents the contribution change between the different algorithms and real trajectories of typescript project in 45 months. The horizontal coordinate shows the change of the month and the vertical coordinate indicates the contribution received for each month.}
\vspace{-0.05cm}  
\setlength{\abovecaptionskip}{-0cm}   
\setlength{\belowcaptionskip}{-0.01cm}   
\label{fig:compare_fig6}
\end{figure}

Minghui Zhou et al. \cite{10} proposed to study the contribution of developers from three aspects: motivation, ability and environment, in which motivation involves various motives, including collaboration, society, compensation, user needs, etc. Minghui Zhou applied the indicators involved to Mozilla and Gnome projects, and then studied the characteristics of long-term developers through machine learning methods. 
Jalerson Lima \cite{11}proposed a set of indicator which were designed to evaluate the developer's contribution, covering four aspects: code contribution, average complexity of the method, introduction of bugs, and bug fixes. Then they obtained empirical evidence from the project and team leaders. \par


\section{Discussion}
We aim to find a way to help developers make decisions that will enhance the contribution of open source projects. The most basic point is whether the definition of the contribution reflects the workload of the developer. Currently, we are only preliminary to consider the number of actions and the effect of each action. Besides, we believe there are many other metrics that can be incorporated that capture a developer's workload. Likewise, consideration of the actions' number needs to involve the more granular content, such as evaluating the impact of how much value the action will bring.\par

\section{Conclusion and Future work}
\;We propose a data-based contribution evaluation model, which can dynamically measure developers' contributions based on changes in data. We address the challenge of how to improve developers' contribution, which is an extremely rare and significant work at present. It is the first time that the presented framework explores deep reinforcement learning techniques to manage open source software, which enables us to design a more robust framework to improve developers' contributions. We have performed extensive experiments, proving the remarkable success of our proposed framework.\par

In future work, we will continue to explore comprehensive and objective measures of developer contribution to meet the needs of actual production. On the other hand, we will also probe how to increase the overall contribution, with collaborating between multiple developers.\par


\bibliographystyle{abbrv}
\bibliography{OSSMentor} 

\begin{thebibliography}{10}

\bibitem{4}
A.~{Bachmann} and A.~{Bernstein}.
\newblock When process data quality affects the number of bugs: Correlations in
  software engineering datasets.
\newblock In {\em 2010 7th IEEE Working Conference on Mining Software
  Repositories (MSR 2010)}, pages 62--71, 2010.

\bibitem{5}
C.~{Bird}, A.~{Gourley}, and P.~{Devanbu}.
\newblock Detecting patch submission and acceptance in oss projects.
\newblock In {\em Fourth International Workshop on Mining Software Repositories
  (MSR'07:ICSE Workshops 2007)}, pages 26--26, 2007.

\bibitem{i8}
B.~{Boehm} and {Li Guo Huang}.
\newblock Value-based software engineering: a case study.
\newblock {\em Computer}, 36(3):33--41, 2003.

\bibitem{i7}
B.~W. Boehm.
\newblock {\em Value-Based Software Engineering: Overview and Agenda}, pages
  3--14.
\newblock Springer Berlin Heidelberg, Berlin, Heidelberg, 2006.

\bibitem{9}
G.~Gousios, E.~Kalliamvakou, and D.~Spinellis.
\newblock Measuring developer contribution from software repository data.
\newblock In {\em Proceedings of the 2008 International Working Conference on
  Mining Software Repositories}, MSR ’08, page 129–132, New York, NY, USA,
  2008. Association for Computing Machinery.

\bibitem{gail}
J.~Ho and S.~Ermon.
\newblock Generative adversarial imitation learning.
\newblock In {\em Advances in neural information processing systems}, pages
  4565--4573, 2016.

\bibitem{6}
C.~{Jensen} and W.~{Scacchi}.
\newblock Role migration and advancement processes in ossd projects: A
  comparative case study.
\newblock In {\em 29th International Conference on Software Engineering
  (ICSE'07)}, pages 364--374, 2007.

\bibitem{i9}
R.~Kazman, R.~Bahsoon, I.~Mistrik, and Y.~Zhang.
\newblock Chapter 1 - economics-driven software architecture: Introduction.
\newblock In I.~Mistrik, R.~Bahsoon, R.~Kazman, and Y.~Zhang, editors, {\em
  Economics-Driven Software Architecture}, pages 1 -- 8. Morgan Kaufmann,
  Boston, 2014.

\bibitem{ddpg}
P.~A. Lillicrap T~P, Hunt J~J.
\newblock Continuous control with deep reinforcement learning.
\newblock {\em arXiv preprint arXiv:1509.02971}, 2015.

\bibitem{11}
J.~{Lima}, C.~{Treude}, F.~F. {Filho}, and U.~{Kulesza}.
\newblock Assessing developer contribution with repository mining-based
  metrics.
\newblock In {\em 2015 IEEE International Conference on Software Maintenance
  and Evolution (ICSME)}, pages 536--540, 2015.

\bibitem{i10}
J.~Marlow, L.~Dabbish, and J.~Herbsleb.
\newblock Impression formation in online peer production: Activity traces and
  personal profiles in github.
\newblock In {\em Proceedings of the 2013 Conference on Computer Supported
  Cooperative Work}, CSCW ’13, page 117–128, New York, NY, USA, 2013.
  Association for Computing Machinery.

\bibitem{3}
S.~K. Sowe, I.~Stamelos, and L.~Angelis.
\newblock Understanding knowledge sharing activities in free/open source
  software projects: An empirical study.
\newblock {\em Journal of Systems and Software}, 81:431--446, 03 2008.

\bibitem{7}
F.~{Thung}, T.~F. {Bissyandé}, D.~{Lo}, and L.~{Jiang}.
\newblock Network structure of social coding in github.
\newblock In {\em 2013 17th European Conference on Software Maintenance and
  Reengineering}, pages 323--326, 2013.

\bibitem{i11}
J.~Tsay, L.~Dabbish, and J.~Herbsleb.
\newblock Influence of social and technical factors for evaluating contribution
  in github.
\newblock In {\em Proceedings of the 36th International Conference on Software
  Engineering}, ICSE 2014, page 356–366, New York, NY, USA, 2014. Association
  for Computing Machinery.

\bibitem{8}
Wikipedia.
\newblock Source lines of code.
\newblock \url{https://en.wikipedia.org/wiki}, 2016.

\bibitem{report}
X-lab.
\newblock Github's digital annual report for 2019.
\newblock In {\em https://github.com/X-lab2017/github-analysis-report-2019},
  2020.

\bibitem{1}
Q.~{Xuan}, M.~{Gharehyazie}, P.~T. {Devanbu}, and V.~{Filkov}.
\newblock Measuring the effect of social communications on individual working
  rhythms: A case study of open source software.
\newblock In {\em 2012 International Conference on Social Informatics}, pages
  78--85, 2012.

\bibitem{2}
Y.~Ye and K.~Kishida.
\newblock Toward an understanding of the motivation open source software
  developers.
\newblock In {\em Proceedings of the 25th International Conference on Software
  Engineering}, ICSE ’03, page 419–429, USA, 2003. IEEE Computer Society.

\bibitem{10}
M.~{Zhou} and A.~{Mockus}.
\newblock Who will stay in the floss community? modeling participant’s
  initial behavior.
\newblock {\em IEEE Transactions on Software Engineering}, 41(1):82--99, 2015.

\end{thebibliography}
\renewcommand\ref{INferencce}

\end{document}